\documentclass[prl,twocolumn,superscriptaddress,showpacs]{revtex4-1}
\usepackage{graphicx}
\usepackage{epsfig}
\usepackage{amssymb,amsmath,amsfonts,hyperref}
\usepackage{wasysym}
\usepackage{latexsym}
\usepackage{verbatim}
\usepackage[normalem]{ulem}
\usepackage{color}

\newcommand{\be}{\begin{eqnarray}}
\newcommand{\ee}{\end{eqnarray}}

\begin{document}
\title{Emergent moments and random singlet physics in a Majorana spin liquid}
\author{Sambuddha Sanyal}
\affiliation{\small{Department of Physics, Indian Institute of Science Education and
Research (IISER) Tirupati, Tirupati 517507, India }}
\author{Kedar Damle}
\affiliation{\small{Department of Theoretical Physics, Tata Institute of Fundamental Research, Mumbai, India}}
\author{J.~T.~Chalker}
\affiliation{\small{Theoretical Physics, University of Oxford, Parks Road OX1 3PU, Oxford, United Kingdom}}
\author{R. Moessner}
\affiliation{\small{Max Planck Institute for the Physics of Complex Systems, N\"{o}thnitzer Str. 38, 01187 Dresden, Germany}}
\begin{abstract}
We exhibit an exactly solvable example  of a SU(2) symmetric Majorana spin liquid phase,  
in which quenched disorder leads to random-singlet phenomenology. 
More precisely, we argue that a strong-disorder fixed point controls the low temperature susceptibility $\chi(T)$ of an exactly solvable $S=1/2$ model on the decorated honeycomb lattice with quenched bond disorder and/or vacancies, leading to $\chi(T) =  {\mathcal C}/T+  {\mathcal D} T^{\alpha(T) - 1}$ where $\alpha(T) \rightarrow 0$ as $T \rightarrow 0$. The first term is a Curie tail that represents the emergent response of vacancy-induced spin textures spread over many unit cells:
it is an intrinsic feature of the site-diluted system, rather than an extraneous effect arising from isolated free spins. The second term, common to both vacancy and bond disorder (with different $\alpha(T)$ in the two cases) is the response of a random singlet phase, familiar from random antiferromagnetic spin chains and the analogous regime in phosphorus-doped silicon (Si:P). 

\end{abstract}


\maketitle
Quantum spin liquids~\cite{Balents_review,Fieldguide,Kitaevreview} are of interest as topological states of magnets. Specifically, their low-energy physics contains unusual degrees of freedom, such as emergent gauge fields and fractionalised excitations carrying  corresponding gauge charge~\cite{Alet_Walczak_Fisher_review}. Although elusive in the ground state of a clean system, these excitations are perhaps the most direct signature of the presence of a spin liquid. As these  need not correspond to known elementary particles, spin liquids can even make quasiparticles with desired quantum numbers  uniquely available \cite{cms}.  
In particular, these excitations can then exhibit interesting, new and unusual low-energy behaviour of their own, where the bulk of the spin liquid can act as a matrix mediating emergent interactions between the defects.

Besides thermal excitations, disorder turns out to be a particularly revealing probe, as it can lead to the appearance (`nucleation') of gauge-charged defects already at $T=0$, and especially for gapless excitations, can rearrange the low-energy spectral weight even when the total amount of disorder is small~\cite{Willans_Chalker_Moessner_PRL,Willans_Chalker_Moessner_PRB,Sen_Damle_Moessner_PRL,Sen_Damle_Moessner_PRB,Rehn_etal,RRPSingh,Kimchi_Nahum_Senthil_PRX,Liu_etal_JQ_PRX,Knolle_Moessner_Perkins}. In reverse, probing the low-energy physics of an experimental compound can provide insights into amount and nature of disorder present in a particular material~\cite{Kimchi_Sheckelton_McQueen_Lee_NatureComm,Kitagawa_etal_H3LiIr2O6,Sheckelton_etal_LiZn2Mo3O8_NatMat,Sheckelton_etal_LiZn2Mo3O8_PRB,Han_etal_herbertsmithite_PRB,Law_etal_TaS2_Proc.Nat.Acad.Sci}.

Our work weaves together several of these threads. We study an SU(2)-invariant version~\cite{Yao_Lee} of Kitaev's model~\cite{Kitaev_anyon}, on a decorated honeycomb lattice, subject to random site (dilution) and bond disorder. This fractionalised magnet exhibits Majorana fermion excitations coupled to an emergent $\mathbb{Z}_2$ gauge field. 

Both types of disorder realise the physics of a strong-disorder random singlet phase, with a characteristic divergence of the low-$T$ susceptiblity, as captured by the second term in 
\begin{eqnarray}
  \chi(T) & = & {\mathcal C}/T+  {\mathcal D} T^{\alpha(T) - 1} \; ,
  \label{eq:susc}
\end{eqnarray}
 with positive $\alpha(T)$ vanishing with $T$. Such a 
random singlet phase is familiar from the physics of random antiferromagnetic spin chains~\cite{Ma_Dasgupta_Hu,Dasgupta_Ma,Bolzer_Gould_Clark,Fisher,Damle_Motrunich_Huse_PRL,Motrunich_Damle_Huse_spinchainPRB} and Si:P~\cite{Bhat_Lee_PRL,Paalanen_Ruckenstein_Thomas,Lee_NMR}, exhibiting
a random pattern of valence bonds between moments, and a broad distribution of triplet excitation energies for the valence bonds. 
A random singlet description~\cite{RRPSingh,Kimchi_Nahum_Senthil_PRX,Liu_etal_JQ_PRX} has been argued to provide a reasonably good phenomenological fit \cite{Kimchi_Sheckelton_McQueen_Lee_NatureComm} to recent experiments on several interesting S = 1/2 magnets with geometric frustration and quenched disorder~\cite{Kitagawa_etal_H3LiIr2O6,Sheckelton_etal_LiZn2Mo3O8_NatMat,Sheckelton_etal_LiZn2Mo3O8_PRB,Han_etal_herbertsmithite_PRB,Law_etal_TaS2_Proc.Nat.Acad.Sci}, provided one includes the effects of spin-orbit coupling whenever present.
Here, we employ the exact solubility of the model combined with arbitrary-precision numerics to pin down the low-$T$ structure in considerable detail. 

Dilution, on top of this random-singlet physics, generates the Curie tail,  ${\mathcal C}/T$ (Eq.~\ref{eq:susc}), in the susceptibility. This is predominantly due to zero modes of the Majorana fermions, rather than isolated spins. Its presence can thus be used to distinguish between the two types of disorder, while the  random singlet signal -- {\it subdominant} in the case of dilution -- exhibits a non-universal exponent depending on the disorder strength.

\begin{figure}
\begin{center}
\includegraphics[width=0.4\columnwidth]{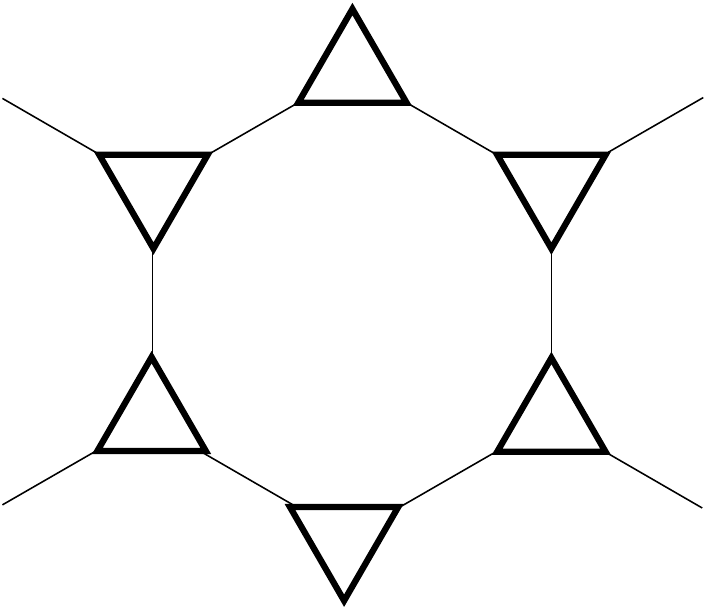}%
\includegraphics[height=0.3\columnwidth,trim = 6.7cm 14.5cm 3cm 4cm, clip]{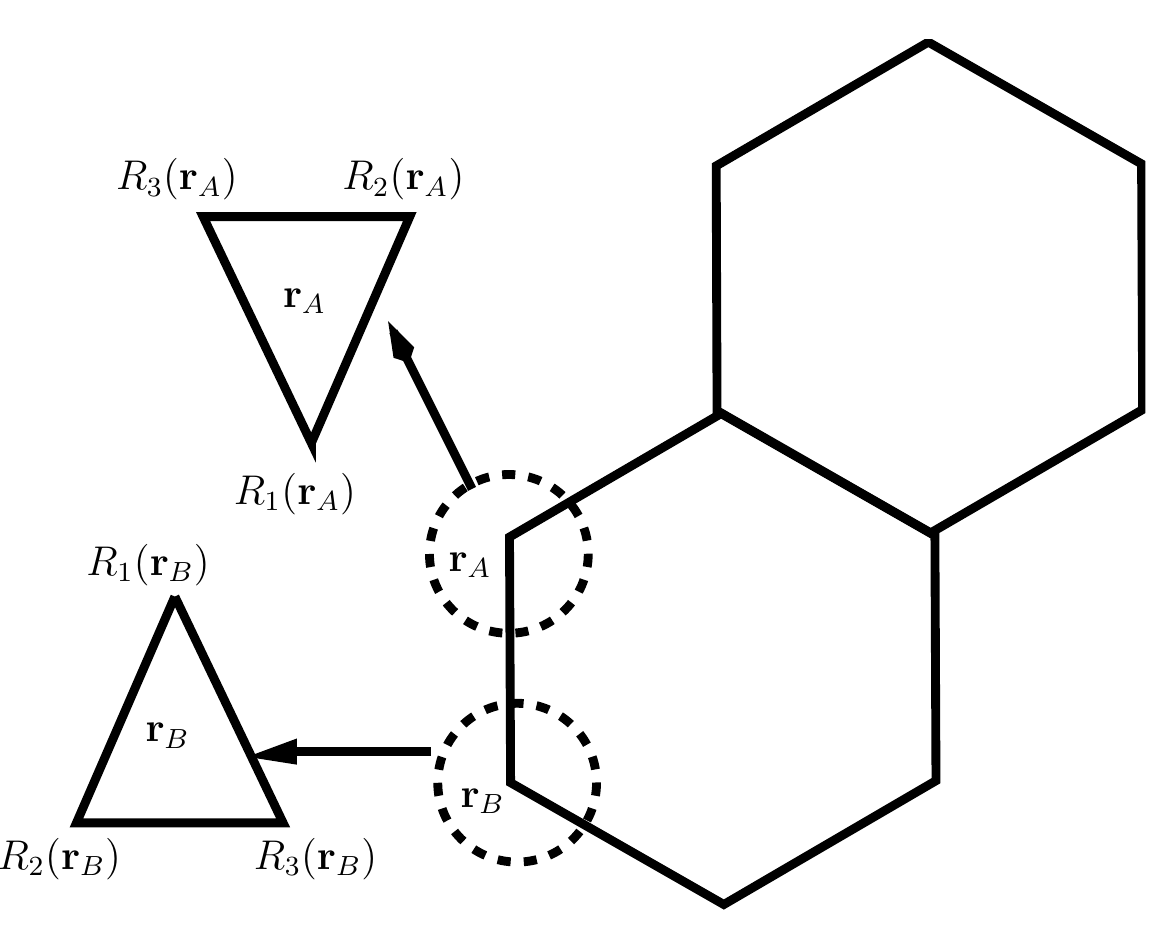}
\caption{The decorated honeycomb, or star, lattice (left) is obtained from the honeycomb on replacing   every  $A$ [$B$] sublattice site $\vec{r}_A$ [$\vec{r}_B$]  of the honeycomb (right) by an up [down] pointing triangle made up of sites $R_{1/2/3}(\vec{r}_A)$ [$R_{1/2/3}(\vec{r}_B)$]}
\label{Fig1template}
\end{center}
\end{figure}

The integrable model~\cite{Yao_Lee} studied here provides an  SU(2) symmetric Majorana spin liquid~\cite{Yao_Lee,Biswas_su2,Lai_Motrunich} in which the $\mathbb{Z}_2$ fluxes are static and gapped. It has $S=1/2$ moments $\vec{S}_R$ on sites $R$ of a decorated honeycomb lattice in which each honeycomb site $\vec{r}$ is replaced by a triangle $\vec{r}$ consisting of three sites $R_{1/2/3}(\vec{r})$ (see Fig.~\ref{Fig1template}). The three spins of a given triangle are coupled to each other with a large antiferromagnetic exchange $J$, the largest energy scale in the Hamiltonian. As a result, the low-energy physics is controlled by states in which each triangle is in one of two $S_{\rm tot} =1/2$ doublet states. These two doublets are distinguished by the ``orbital'' quantum number $\tau^z_{\vec{r}} = \pm 1$. Neighbouring triangles are coupled by a multi-spin interaction of strength $K$ that is sensitive to the ``orbital wavefunction'' of the total spin state of each triangle.

When $K \ll J$, the low-energy Hamiltonian can be written in terms of spin-half variables $\vec{S}_{\vec{r}} = \vec{\sigma}_{\vec{r}}/2$ (where $\vec{\sigma}_{\vec{r}}$ are Pauli-matrices) representing the total spin of each triangle:
\begin{equation}
{\mathcal H}_{\rm YL} = \frac{1}{2}\sum_{\langle \vec{r} \vec{r}^{'} \rangle_{\lambda}} K \tau_{\vec{r}}^{\lambda} \tau_{\vec{r}^{'}}^{\lambda} \vec{\sigma}_{\vec{r}} \cdot \vec{\sigma}_{\vec{r}^{'}} -\sum_{\vec{r}} \vec{B} \cdot \vec{S}_{\vec{r}} \; .
\end{equation}
Here, the first sum is over all nearest-neighbour links $\langle \vec{r} \vec{r}^{'}\rangle_\lambda$ on the honeycomb lattice, $\lambda=x,y,z$ denotes the orientation
of the nearest neighbour link connecting the $A$-sublattice site $\vec{r}$ to the $B$-sublattice
site $\vec{r}^{'}$ of the honeycomb lattice, $\tau^z_{\vec{r}}$ is the orbital quantum number introduced above, and $\tau^{x,y,z}_{\vec{r}}$ 
are Pauli matrices in the orbital Hilbert space at each $\vec{r}$.

Nonmagnetic impurities, corresponding to missing spins in the original model on the decorated honeycomb lattice, are represented by missing sites in this low-energy Hamiltonian since a single vacancy on triangle $\vec{r}$ leads to a nondegenerate (inert) singlet state for this triangle,  $\vec{S}_{\vec{r}} =0$. In other words, a nonmagnetic impurity on triangle $\vec{r}$ of the original model leads to the corresponding site being deleted in ${\mathcal H}_{YL}$ (along with the three $K$-bonds connecting it to neighbours on the honeycomb lattice). In addition, bond disorder in the coupling $K$ of ${\mathcal H}_{YL}$ arises as a consequence of quenched disorder in the strength of the multispin interaction, so long as the intra-triangle exchange $J$ remains the largest coupling in the system.
In what follows, we will assume this to be true and analyze the effects of bond and site disorder in ${\mathcal H}_{YL}$.

We begin by noting that the Hamiltonian ${\mathcal H}_{YL}$ admits an exact solution~\cite{Yao_Lee,Kitaev_anyon} in terms
of a Majorana representation~\cite{Tsvelik,Shastry_Sen}:
$\sigma^{z}_{\vec{r}} = -ic^{x}_{\vec{r}} c^{y}_{\vec{r}}$, $\tau^{z}_{\vec{r}} = -ib^{x}_{\vec{r}} b^{y}_{\vec{r}}$, and cyclic permutations thereof, where $c^{\lambda}_{\vec{r}}$ and $b^{\lambda}_{\vec{r}}$ are Majorana (real) fermion
operators. In the physical Hilbert space, which is characterized by the local constraint:
$D_{\vec{r}} \equiv c^x_{\vec{r}}c^y_{\vec{r}}c^z_{\vec{r}}b^x_{\vec{r}}b^y_{\vec{r}}b^z_{\vec{r}}= i$, we have the identitiy $\sigma^{\alpha}_{\vec{r}} \tau^{\beta}_{\vec{r}} = ic^{\alpha}_{\vec{r}} b^{\beta}_{\vec{r}}$. Defining $\mathbb{Z}_2$ gauge fields 
$u_{\vec{r}\vec{r}^\prime}\equiv b^\lambda_{\vec{r}}b^\lambda_{\vec{r}^\prime}$ on $\lambda$-links, 
the problem reduces in these variables to three flavours of $c$ fermions hopping on the honeycomb lattice while coupled to a common $\mathbb{Z}_2$ gauge field $u$ that has no quantum dynamics:
\begin{equation}
{\mathcal H}_{\rm YL} =  \frac{J}{4}\sum_{\alpha=x,y,z}\sum_{\vec{r} \vec{r}^{'} }iu_{\vec{r} \vec{r}^{'} }c^{\alpha}_{\vec{r}} c^{\alpha}_{\vec{r}^{'}} + \frac{h}{2}\sum_{\vec{r}} i c^{x}_{\vec{r}} c^{y}_{\vec{r}} \; ,
\end{equation}
where $\vec{B} = h\hat{z}$.  In other words, the model reduces to a collection of free fermion problems, one for each static configuration of $\mathbb{Z}_2$ fluxes threading faces of the lattice. In consequence, the temperature-dependent susceptibility for the spin model, including the effects of exchange disorder and vacancies, can be determined from the density of states of an associated free fermion system. 
\begin{figure}
\begin{center}
\includegraphics[width=0.75\columnwidth]{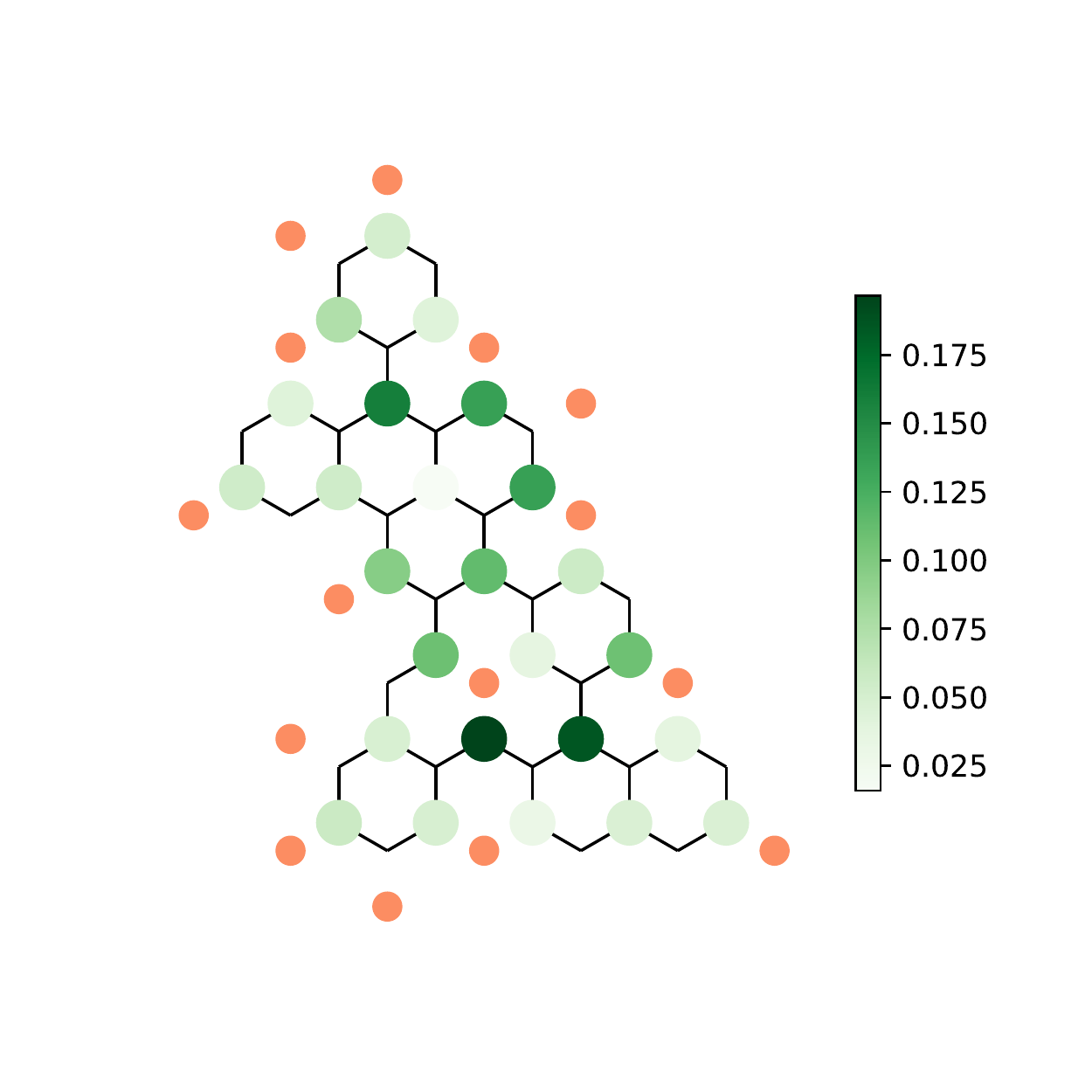}
\caption{Spin polarization $\langle \sigma^z_r \rangle$ induced by an infinitesimal $z-$magnetic field at $T=0$. This reflects two disorder-robust Majorana zero modes of the Majorana Hamiltonian with $\pi$ flux attached to each vacancy, and with exchange couplings drawn from a uniform
distribution whose half-width is $20\%$ of the mean exchange.
This local moment is nonzero only on one sublattice of sites, and strictly localized. Orange circles denote vacancies in  and immediately adjacent to this region; bonds  to the outside of the region are not shown.}
\label{Figwavefunctiontemplate}
\end{center}
\end{figure}

A dictionary relating the free fermion problem to the magnetic one is as follows. Physical properties of the SU(2) symmetric model are controlled by the behaviour of a triplet of Majorana fermions hopping on the honeycomb lattice. The hopping matrix ${\mathbf K}$ has matrix elements $\pm i K_{\vec{r}_A \vec{r}_B} u_{ \vec{r}_A \vec{r}_B}$ with the $u$ corresponding to the ground state flux sector (here $\vec{r}_A$ and $\vec{r}_B$ are the $A$ and $B$ sublattice sites connected by the corresponding link of the honeycomb lattice). We introduce  canonical (complex) fermions, defined as $f_{\vec{r}} = \pm(c^{x}_{\vec{r}} + i c^{y}_{\vec{r}})/2$, with the plus (minus) sign on $A$ ($B$) sublattice sites. The $f$-fermion Hamiltonian is then a tight-binding model with hopping matrix ${\mathbf K}$, single-particle eigenenergies $\epsilon$, and density of states $\rho(\epsilon)$. The magnetic field $h$ acts as a chemical potential for the $f$-fermions, since $S^z_{\vec{r}} = -ic^{x}_{\vec{r}} c^{y}_{\vec{r}} = 1/2-f^{\dagger}_{\vec{r}} f_{\vec{r}}$. The key conclusion is hence that the magnetic susceptibility of the spin model is equivalent to the compressibility of the fermion system and determined by $\rho(\epsilon)$. This conclusion must be augmented with a discussion of the consequences of the constraint $D_{\vec{r}}$ and of the flux sector selected at low temperature~\cite{Kitaev_anyon,Lieb,Zschocke_Vojta,Pedrocchi_Chesi_Loss}: 
we omit details for brevity. 

\begin{figure}
\begin{center}
\includegraphics[width=0.9\columnwidth]{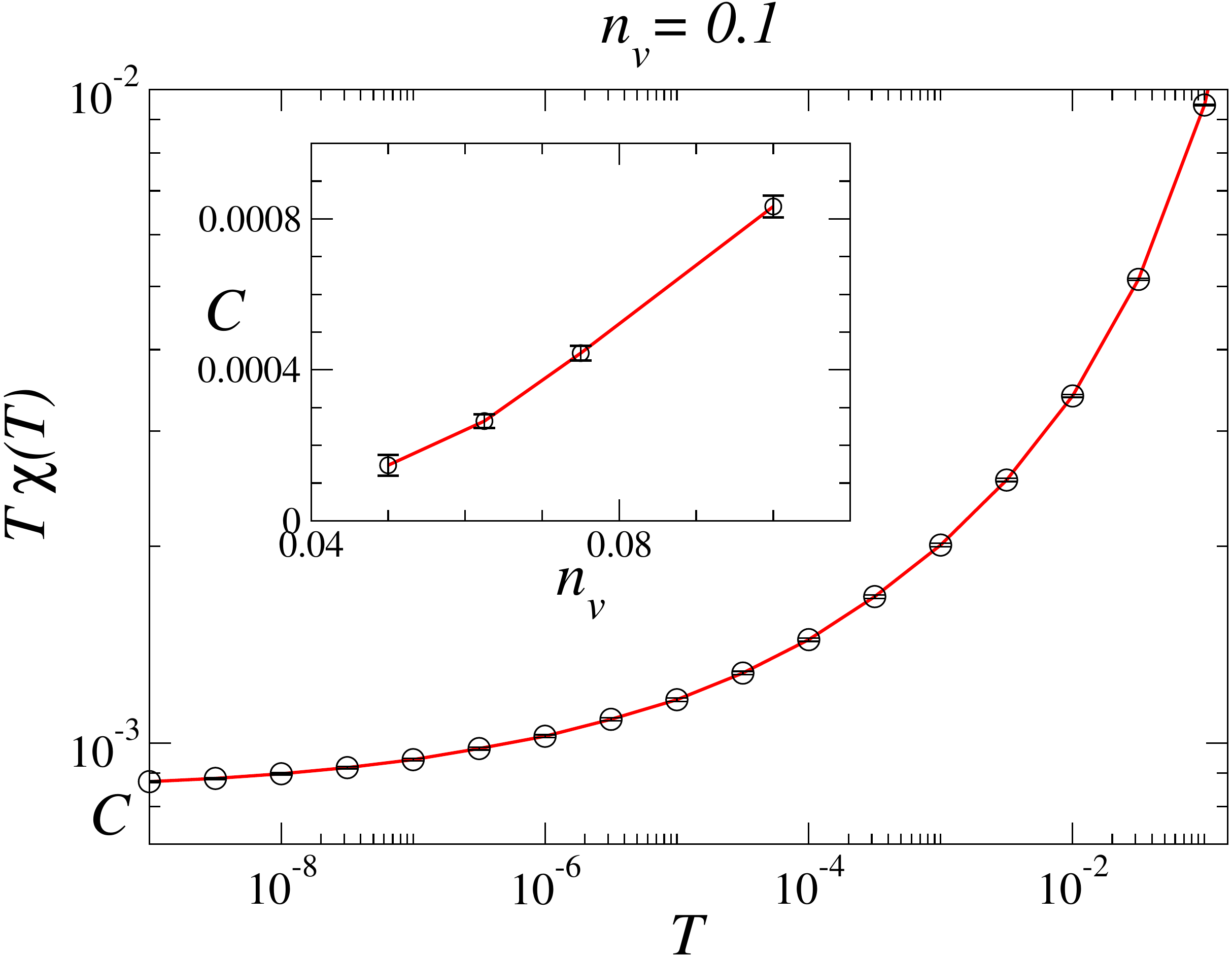}\\
\includegraphics[width=0.9\columnwidth]{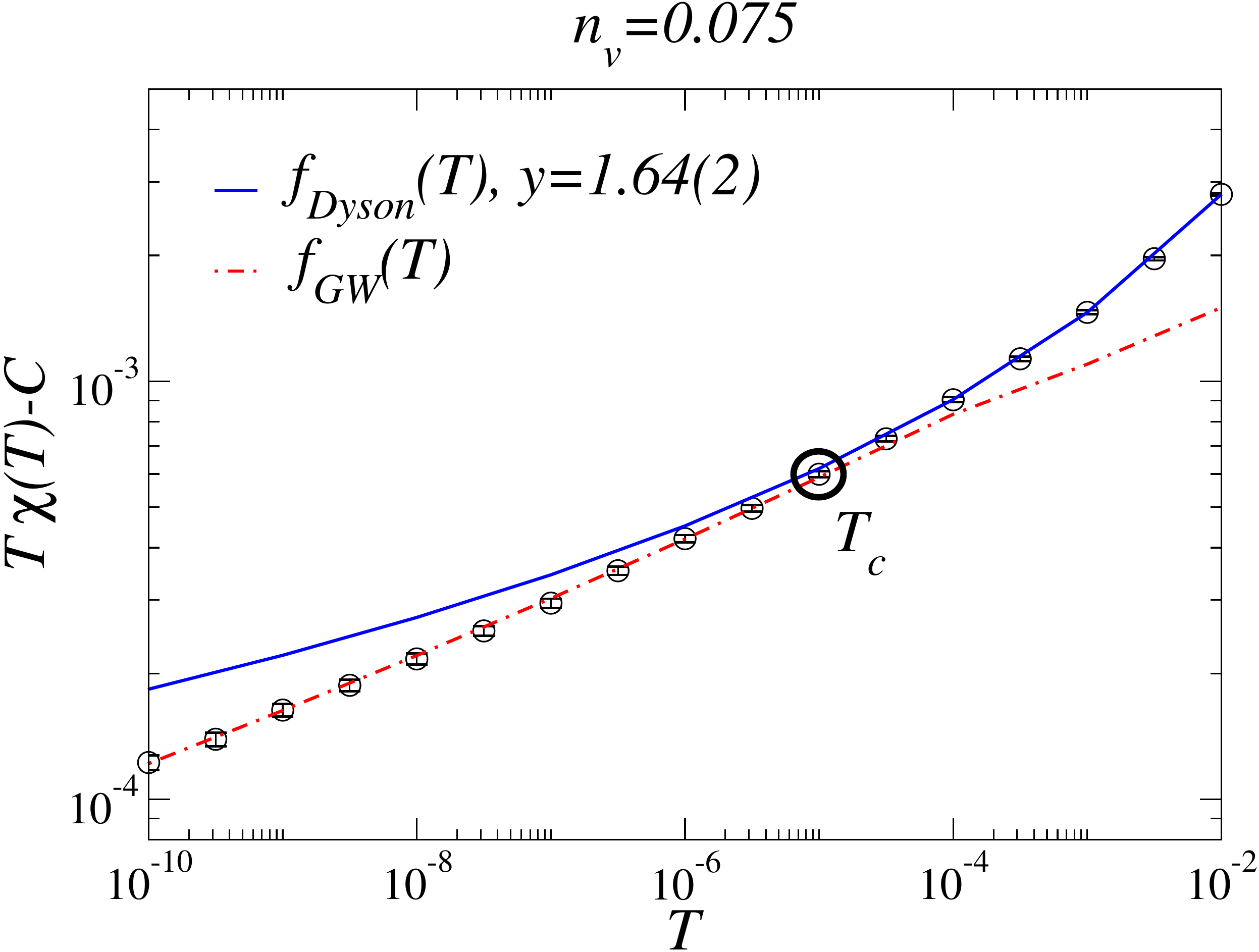}\\
\includegraphics[width=0.49\columnwidth]{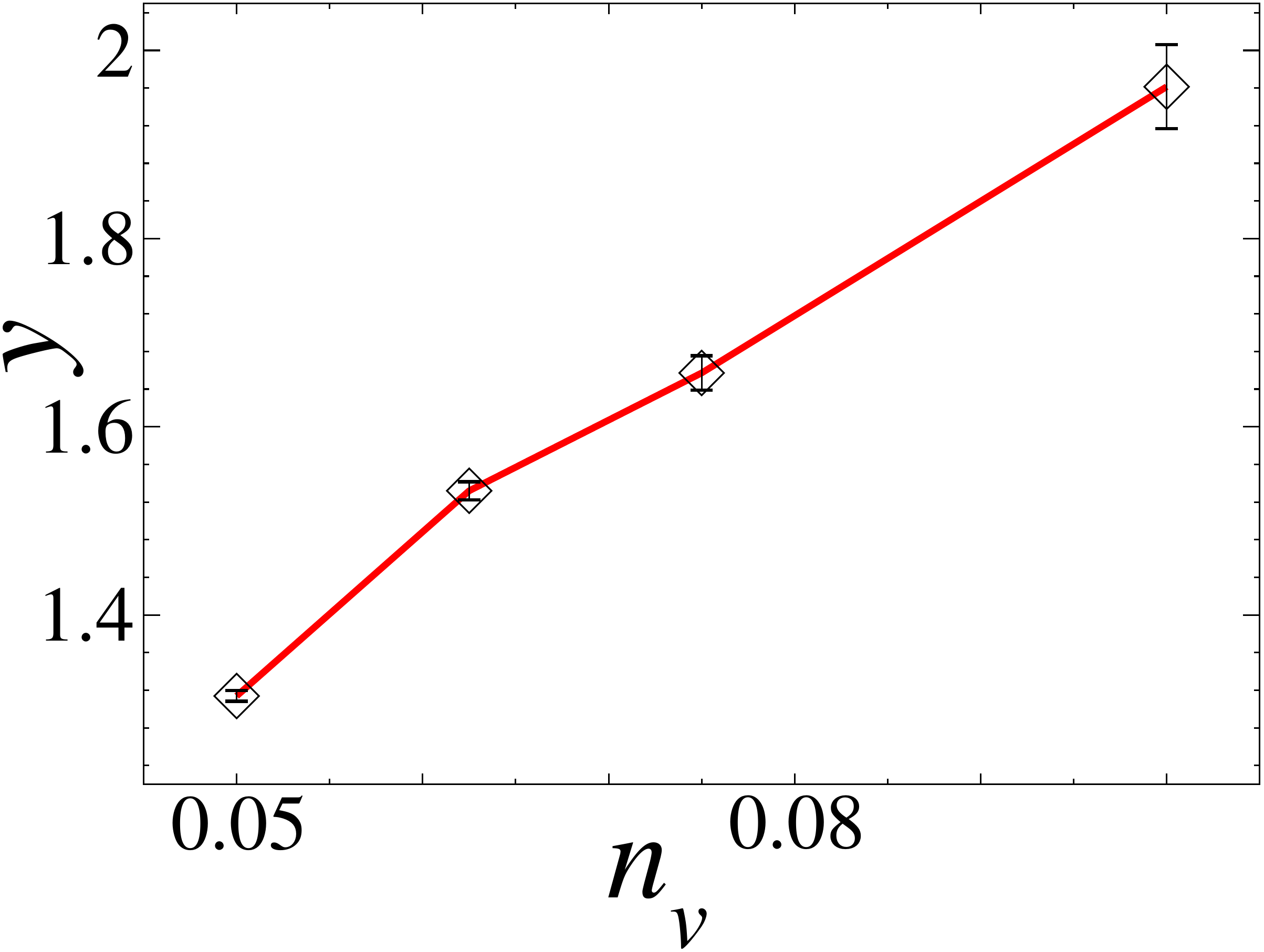}
\includegraphics[width=0.49\columnwidth]{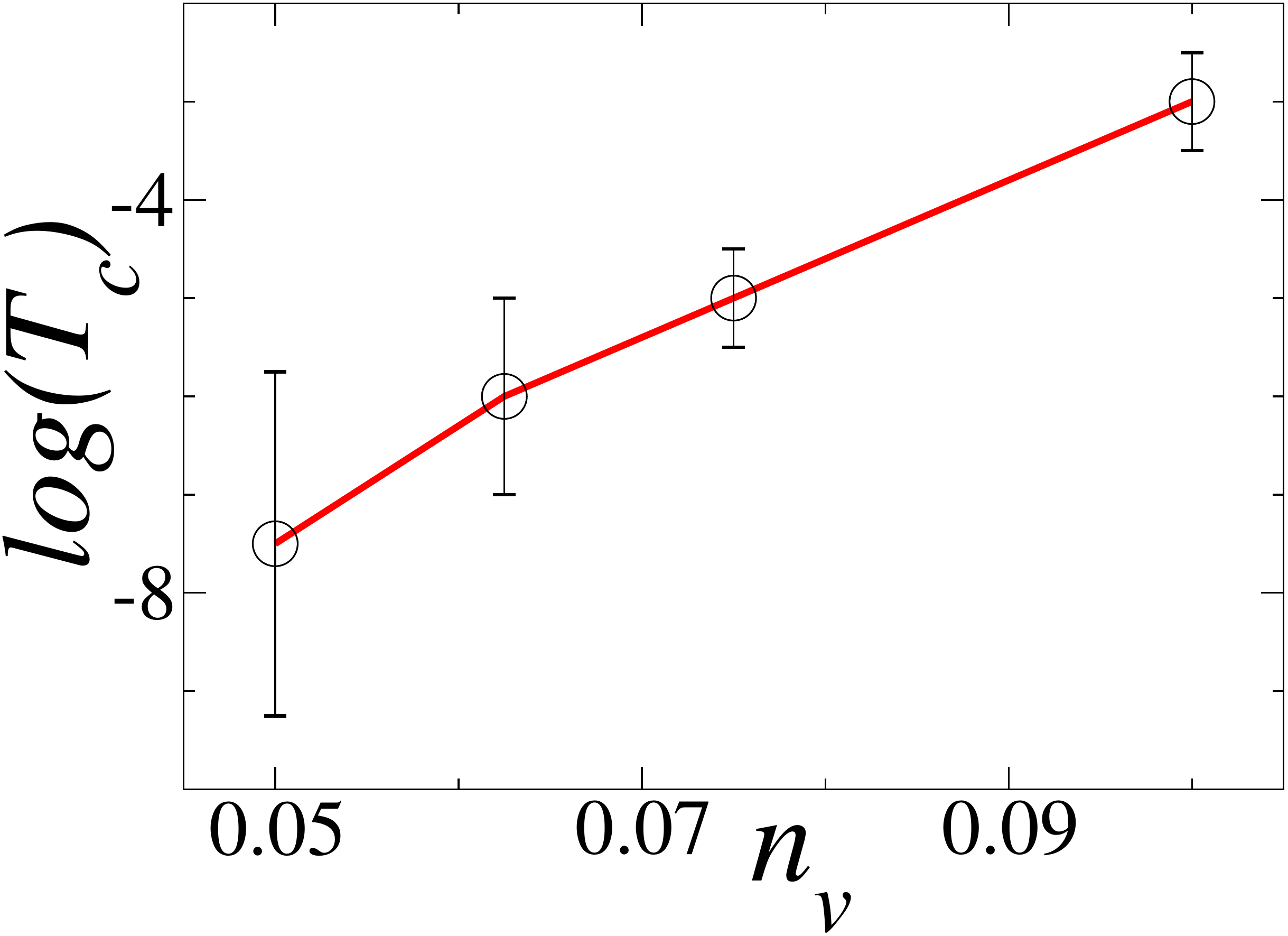}
\caption{Top panel: $T \chi(T)$ for a system with density of vacancies $n_v=6.25\%$; 
note the saturation at low temperature, corresponding to a nonzero Curie constant ${\mathcal C}$. Inset shows the $n_v$ dependence of ${\mathcal C}$. Middle panel: $T\chi(T) -{\mathcal C}$ shows a clear crossover at a temperature scale $T_c$ from the functional form $f_{\mathrm{Dyson}} = a/(\log^y(K_{\mathrm{av}}/T))$ at intermediate temperature to  $f_{\mathrm{GW}} = b \exp({-c\log^{2/3}(K_{\mathrm{av}}/T)})$ at the lowest temperatures. Bottom panel: Dependence on $n_v$ of $y$ (left) and $T_c$ (right).
}
\label{Fig2template}
\end{center}
\end{figure}

In general, chiral symmetry of ${\mathbf K}$ ensures that $\rho(\epsilon)$ is an even function of $\epsilon$, and for the clean system $\rho(\epsilon) \propto |\epsilon|$. Disorder affects $\rho(\epsilon)$ in two possible ways. Exchange randomness generates additional low-energy fermion states with a continuous distribution of energies and a density $\rho_{\rm reg}(\epsilon)$ that diverges as $\epsilon\to 0$. Site dilution produces both a continuous contribution $\rho_{\rm reg}(\epsilon)$ (similarly divergent for $\epsilon\to 0$, albeit with a different functional form) and also a finite density of zero modes, so that in this case $\rho(\epsilon) = \rho_0\delta(\epsilon) + \rho_{{\mathrm{reg}}}(\epsilon)$. 
It is convenient to measure fermion energy and temperature in units of the disorder-averaged exchange coupling $K_{\rm av}$ and to use the parameterisations $\Gamma_\epsilon\equiv \log_{10}(K_{\rm av}/\epsilon)$ and $\Gamma_T \equiv \log_{10}(K_{\rm av}/T)$. In addition, we express the integrated density of states in terms of $\Gamma_\epsilon$ via $N(\Gamma_\epsilon) =\int_{-\epsilon}^{\epsilon}d\epsilon^\prime \rho_{\rm reg}(\epsilon^\prime)$. 

An elementary calculation relates the susceptibility to the compressibility of the canonical fermions. For $T \ll K_{\mathrm{av}}$ we obtain
\begin{eqnarray}
\chi(T) &=& \frac{\rho_0}{4T} + \frac{N(\Gamma_T)}{4T} \; .
\label{eq:susc2}
\end{eqnarray}
At low temperature, $\Gamma_T$ is large and $N(\Gamma_T)$ reflects the form of $\rho_{\rm reg}(\epsilon)$ near $\epsilon=0$.

Consequences of this mapping for magnetic properties are summarised in Fig.~\ref{Fig2template}. Zero modes of the fermion system yield a Curie tail in the susceptibility, with coefficient ${\cal C} = \rho_0/4$. That this is a cooperative response and not the result of individual free spins is reflected in the form of the corresponding fermion eigenfuctions, which extend over multiple sites.  Disorder-induced low-energy fermion states with density $\rho_{\rm reg}(\epsilon)$ are responsible for a second contribution to $\chi(T)$, also divergent for $T\to 0$ but subleading.

These statements can be backed up in considerable detail by appealing to a mapping of the spin liquid to a random bipartite hopping problem~\cite{Gade,Gade_Wegner,Motrunich_Damle_Huse_dirtySCPRB,Motrunich_Damle_Huse_GadeWegnerPRB,Mudry_Ryu_Furusaki_PRB}, from which results can be transposed into the present context. Justification of the framework
appeals to an underlying infinite-disorder fixed point as  the origin of the  random-singlet physics. The $T$-dependence of the second term in Eq.~\ref{eq:susc2} turns out to be   $N(\Gamma_T)/T=T^{\alpha(T) -1}$,  with $\alpha(T)$ vanishing slowly but non-universally as $T \rightarrow 0$, so that different $\alpha(T)$ arise for site and bond disorder. 

In the case of bond disorder, the form of $\alpha(T)$ follows from the fact that  for the random hopping problem  $|dN(\Gamma)/d\Gamma|$ for large $\Gamma$ has the modified Gade-Wegner form~\cite{Gade,Gade_Wegner} $a\exp(-b \Gamma^{1/x})$, with $x=3/2$ ~\cite{Motrunich_Damle_Huse_GadeWegnerPRB,Mudry_Ryu_Furusaki_PRB}. For the drifting susceptibility exponent $\alpha(T)$, this implies vanishing $\alpha(T)  \sim 1/\Gamma_T^{1/3}$ in the low temperature limit~\cite{footnote}. Thus, bond disorder on its own~\cite{Motrunich_Damle_Huse_GadeWegnerPRB}
provides a low temperature susceptibility of purely random-singlet form.

For site dilution, our numerical results are summarized below. 
We find $\alpha(T) \sim y(n_v) \log(\Gamma_T)/\Gamma_T$ at not-too-low temperature, which crosses over to $\alpha(T) \sim 1/\Gamma_T^{1/3}$ below the crossover temperature $T_c \sim K_{\mathrm{av}} 10^{-\Gamma_c(n_v)}$. Interestingly, for the site-diluted case, $y(n_v)$ and $T_c(n_v)$ both decrease quite rapidly with decreasing concentration $n_v$ of vacancies, implying a correspondingly stronger singularity in the random singlet form of the susceptibility for lower values of vacancy density. We note that, 
as all the key features of the low-energy physics of the random bond-disordered case are already present here, 
further adding such  disorder is not expected to lead to any qualitative changes.


The results summarised above are supported by extensive numerical calculations, as illustrated in Fig.~\ref{Fig2template}. 
Crucially, these calculations use the methods of Ref.~\cite{Sanyal_Damle_Motrunich_PRL} to extend to much lower energies previous studies \cite{Willans_Chalker_Moessner_PRL,Willans_Chalker_Moessner_PRB}, which were limited to systems with vacancies but no exchange disorder. High resolution in the low-energy density of states is essential to identify the behaviour described above for $N(\Gamma)$ and to distinguish exact zero modes from very low-energy contributions to $\rho_{\rm av}(\epsilon)$. The  form of $\mathbf{K}$ relevant to low-temperature properties of the spin system with site dilution has a $\mathbb{Z}_2$ flux bound to every vacancy, and differs in this way from the tight-binding models for disordered graphene studied in Ref.~\cite{Sanyal_Damle_Motrunich_PRL,Hafner_etal_PRL,Ostrovsky_etal_PRL,Weik_etal_PRB}. Interestingly, the main features of the low-energy density of states are common to systems with and without $\mathbb{Z}_2$ fluxes.

To obtain insight into the physical origin of the Curie contribution to Eq.~(\ref{eq:susc2}), which arises in systems with vacancies, it is necessary to probe in detail the nature of zero modes in the fermion description. To this end it is useful to compare behaviour in systems with different numbers of vacancies, and with or without 
disorder in the hopping amplitudes. A single vacancy necessarily gives rise to a single zero mode, because it creates an imbalance sublattice site numbers for the bipartite hopping problem. In contrast, zero modes at finite vacancy density are a complicated multivacancy effect, spread over many unit cells. As noted in Ref.~\cite{Sanyal_Damle_Motrunich_PRL}, the density of these vacancy-induced zero modes has contributions from both ``fragile'' zero modes (which are sensitive to the values of the hopping matrix elements) and generic ``disorder-robust'' zero modes, which remain pinned to zero energy independent of the disorder in the hopping amplitudes.  

Fig.~\ref{Figwavefunctiontemplate} visualises such a set of local zero modes,
obtained using the methods of Ref.~\onlinecite{Biswas_etal}. 
Based on a comparison of the $\rho_0$ computed here with the corresponding results of Ref.~\cite{Sanyal_Damle_Motrunich_PRL} without flux attachment, we conclude that such disorder-robust zero modes dominate over fragile ones. Thus, we expect the Curie constant ${\mathcal C}$ to be largely determined by the vacancy density and relatively insensitive to bond disorder.
 
These results fit snugly into the random singlet phenomenology advocated~\cite{Kimchi_Sheckelton_McQueen_Lee_NatureComm}  as a reasonably good description of the low temperature physics of a variety of  interesting $S=1/2$ magnets with geometric frustration and quenched disorder
~\cite{Kitagawa_etal_H3LiIr2O6,Sheckelton_etal_LiZn2Mo3O8_NatMat,Sheckelton_etal_LiZn2Mo3O8_PRB,Han_etal_herbertsmithite_PRB,Law_etal_TaS2_Proc.Nat.Acad.Sci}, 
with due allowance for spin-orbit coupling as necessary. 
From this perspective, our work adds two interesting new angles. Firstly, it transfers the exact solubility of Kitaev models to  the SU(2) random singlet case, thereby permitting us to make controlled statements about a vast temperature range for unusually large systems, compared to what is commonly possible for disordered quantum magnets. Second, it adds a Curie tail to the random singlet physics, with the tail arising from spatially extended vacancy-induced spin textures, rather than the response of isolated free spins. This in turn extends a similar phenomenon from the classical realm, known there as orphan spins \cite{schifferdaruka,berlinsky}, whose behaviour in a quantum context has  remained a puzzle.  

Given the rather comprehensive theoretical understanding we have described, the question of experimental implications naturally follows. 
First of all, the insensitivity of our results with regard to the choice of flux sector -- the phenomenology is explicitly the same with and without the bound fluxes -- implies that perturbations the only role of which is to  favour a different ground-state flux configuration will likely not affect the basic phenomenology. In addition, this may provide some robustness against endowing the  fluxes with dynamics of their own, as happens when one tunes away from the exactly soluble point. 

For any such perturbations, one would as is customary for gapless frustrated magnets, expect a crossover to different 
behaviour at an energy scale set by the strength of the perturbation. We emphasize that two central aspects of our preceding discussion should be robust: firstly, the response above this scale; and secondly, the integrated spectral weight below this scale, comprised of the random singlet density of states and that of the Curie tail. 

What happens below this scale is then an interesting many-body problem of its own, which a priori needs to be addressed on a case-by-case basis. For appropriately chosen, or perhaps even naturally occurring, peturbations, this may even yield as yet unexplored variants of cooperative physics of the low-energy emergent degrees of freedom in a disordered, strongly interacting topological magnet.

Finally, on the technical side,  we note that our work also suggests interesting questions for future work. Can one construct a strong-disorder RG procedure directly in spin language for this exactly solvable model? How are emergent moments seeded in this description, and what prevents them from being quenched by singlet valence bonds? And finally, can this RG approach be used to perturb away from the solvable model?

\textit{Acknowledgements:} We thank F. Evers, J. Knolle,  O.~I.~Motrunich, N. Perkins and M. Vojta for engaging discussions, and R.~Bhola and S.~Biswas for assistance with Fig.~\ref{Figwavefunctiontemplate}. KD gratefully acknowledges a fruitful
collaboration\cite{Biswas_etal} with R.~Bhola, S.~Biswas, and M.~Islam on the random geometry of disorder-robust zero modes.
SS and KD gratefully acknowledge use of departmental computational
resources of DTP-TIFR and ICTS-TIFR for all the computational work described here. The computational study described here formed part of the Ph.D thesis work of SS, funded by a fellowship from the TIFR. KD gratefully acknowledges partial support from the
Infosys-Chandrasekharan Center for Random Geometry at the TIFR. JTC thanks the Dept.\ of Theoretical Physics of the TIFR for hospitality during a visit in which this collaboration was initiated, and gratefully acknowledges travel support from the Oxford-India Theoretical Physics Network for this visit; his work is also supported by EPSRC Grant No EP/S020527/1.  This work was in part supported by the Deutsche Forschungsgemeinschaft  under grants SFB 1143 (project-id 247310070) and the cluster of excellence ct.qmat (EXC 2147, project-id 390858490).


\begin{thebibliography}{10}
\bibitem{footnote} Very similar results have been obtained more recently for the spectrum averaged over flux sectors~\protect{\cite{Knolle_Moessner_Perkins}}, consistent with the expectation~\protect{\cite{Motrunich_Damle_Huse_GadeWegnerPRB,Mudry_Ryu_Furusaki_PRB,Gade,Gade_Wegner}} that the low-energy behaviour would be independent of flux sector.
\end{thebibliography}
\end{document}